# Effects of pyruvate administration on infarct volume and neurological deficits following permanent focal cerebral ischemia in rats

Authors: Armando González-Falcón [‡], Eduardo Candelario-Jalil [‡]*, Michel García-Cabrera and Olga Sonia León

Affiliation: Department of Pharmacology, University of Havana (CIEB-IFAL), Apartado Postal 6079, Havana City 10600, Cuba.

[‡] These authors contributed equally to this report.

*Author to whom all correspondence should be addressed:

**Eduardo Candelario-Jalil, Ph.D.**
**Department of Pharmacology**
**University of Havana (CIEB-IFAL)**
**Apartado Postal 6079**
**Havana City 10600**
**CUBA**
**Tel.: +53-7-271-9534**
**Fax: +53-7-336-811**
**E-mail: candelariojalil@yahoo.com**

**Acknowledgements:** The authors are greatly indebted to Dr. Stefano L. Sensi (Department of Neurology, University of California, Irvine, CA, USA) for his critical comments on the manuscript.

**ABSTRACT**

Recent experimental evidences indicate that pyruvate, the final metabolite of glycolysis, has a remarkable protective effect against different types of brain injury. The purpose of this study was to assess the neuroprotective effect and the neurological outcome after pyruvate administration in a model of ischemic stroke induced by permanent middle cerebral artery occlusion (pMCAO) in rats. Three doses of pyruvate (250, 500 and 1000 mg/kg; i.p.) or vehicle were administered intraperitoneally 30 min after pMCAO. In other set of experiments, pyruvate was given either before, immediately after ischemia or in a long-term administration paradigm. Functional outcome, mortality and infarct volume were determined 24 h after stroke. Even when the lowest doses of pyruvate reduced mortality and neurological deficits, no concomitant reduction in infarct volume was observed. The highest dose of pyruvate increased cortical infarction by 27 % when administered 30 min after pMCAO. In addition, when pyruvate was given before pMCAO, a significant increase in neurological deficits was noticed. Surprisingly, on the contrary of what was found in the case of transient global ischemia, present findings do not support a great neuroprotective role for pyruvate in permanent focal cerebral ischemia, suggesting two distinct mechanisms involved in the effects of this glycolytic metabolite in the ischemic brain.

**1. INTRODUCTION**
Stroke is the second most frequent cause of death and only heart disease causes higher mortality. Stroke is most commonly the result of an obstruction of blood flow in a major cerebral vessel (e.g., the middle cerebral artery), which, if not resolved within a short period of time, will lead to an infarcted tissue that may not be therapeutically salvaged [16,35,36].

Development of an effective therapeutic strategy for stroke has been a priority of neuroscientists for decades. Although some clinical benefits have been obtained with the antioxidants ebselen and edaravone [11,39], no neuroprotective agents has been shown conclusively to be clinically effective to prevent or restrict acute neuronal damage after stroke [6,12].

Ischemia-induced neuronal loss is associated with numerous biochemical events initially triggered by the extracellular accumulation of glutamate. In turn, excitotoxicity leads to membrane depolarization, increased concentrations of intracellular calcium, overproduction of reactive oxygen species, inflammation and activation of apoptotic pathways contributing to the progression of tissue damage [2,4,5,7,25,38]. Recent evidences indicate that, in addition to calcium, endogenous zinc may play a role as an ionic mediator of neuronal death, activating various cell death cascades, such as free radical generation and caspase activation [3,21,26,41,48,53,56].

Interestingly, pyruvate, the end metabolite of the glycolytic pathway, protects striatal neurons against excitotoxicity induced by a 30-min exposure to N-methyl-D-aspartate [29,47], prevents neuronal death induced by exogenous and endogenous $H_2O_2$ in cultured neurons [8,30,31], protects almost completely against zinc neurotoxicity [23,49] and prevents $H_2O_2$-induced apoptosis [43]. In addition, results from a very recent report indicate that administration of pyruvate provides spectacular protection against hippocampal CA1 neuronal injury following transient global cerebral ischemia in rats [23].

In the light of all these evidences, the present study was conducted to assess whether pyruvate would show neuroprotective efficacy on the cerebral infarction induced by permanent middle cerebral artery occlusion (pMCAO), a clinically-relevant model of stroke.

**2. METHODS**

*2.1. Animals*
Male Sprague-Dawley rats (CENPALAB, Havana, Cuba) weighing 280-340 g at the time of surgery were used in the present study. Our institutional animal care and use committee approved the experimental protocol (No. 03/12). The animals were quarantined for at least 7 days before the experiment. Animals were housed in groups in a room whose environment was maintained at 21-25 ºC, 45-50 % humidity and 12-h light/dark cycle. They had free access to pellet chow and water. Animal housing, care, and application of experimental procedures were in accordance with institutional guidelines under approved protocols.

*2.2. Permanent focal ischemia model*
Rats were anesthetized with chloral hydrate (300 mg/kg body weight, i.p.). Once surgical levels of anesthesia were attained (assessed by absence of hind leg withdrawal to pinch), ischemia was induced by using an occluding intraluminal suture [9,22,24]. Briefly, the right common carotid artery (CCA) was exposed by a ventral midline neck incision and ligated with a 3-0 silk suture. The pterygopalatine branch of the internal carotid artery was clipped to prevent incorrect insertion of the occluder filament. Arteriotomy was performed in the CCA approximately 3 mm proximal to the bifurcation and a 3-0 monofilament nylon suture, whose tip had been rounded by being heated near a flame was introduced into the internal carotid artery (ICA) until a mild resistance was felt (18-19 mm). Mild resistance to this advancement indicated that the intraluminal occluder had entered the anterior cerebral artery and occluded the origin of the anterior cerebral artery, the middle cerebral artery (MCA) and posterior communicating arteries [22]. After the advancement of the nylon suture, the ICA was firmly ligated with a 3-0 silk suture. The incision was closed and the occluding suture was left in place until sacrificing the animals. The animals were allowed to recover from anesthesia on an electrical heated blanket and to eat and drink freely.

To allow for better postoperative recovery, we chose not to monitor physiological parameters in the present study because additional surgical procedures are needed for this monitoring. Nevertheless, we performed a separate experiment to investigate the effects of pyruvate on major physiological variables in ischemic rats (see Results section).

### 2.3. Neurological Evaluation

Neurological evaluations were performed according to a six-point scale: 0= no neurological deficits, 1= failure to extend left forepaw fully, 2= circling to the left, 3= falling to left, 4= no spontaneous walking with a depressed level of consciousness, 5= death [27,32]. The investigator performing the neurological evaluation did not know the identity of the experimental groups until completion of data analysis.

### 2.4. Quantification of brain infarct volume

The method for quantification of infarct volume was performed exactly as previously reported [50,55]. Briefly, after completing the neurological evaluation at 24 h after permanent focal cerebral ischemia, the animals were sacrificed under deep anesthesia and brains were removed, frozen and coronally sectioned into six 2-mm-thick slices (from rostral to caudal, first to sixth). The brain slices were incubated for 30 min in a 2% solution of 2,3,5-triphenyltetrazolium chloride (TTC) (Sigma Chemical Co.) at 37 °C and fixed by immersion in a 10% phosphate-buffered formalin solution. Six TTC-stained brain sections per animal were placed directly on the scanning screen of a color flatbed scanner (Hewlett Packard HP Scanjet 5370 C) within 7 days. Following image acquisition, the image were analyzed blindly using a commercial image processing software program (Photoshop, version 7.0, Adobe Systems; Mountain View, CA). Measurements were made by manually outlining the margins of infarcted areas. The unstained area of the fixed brain section was defined as infarcted. Cortical and subcortical uncorrected infarcted areas and total hemispheric areas were calculated separately for each coronal slices. Total cortical and subcortical uncorrected infarct volumes were calculated by multiplying the infarcted area by the slice thickness and summing the volume of the six slices. A corrected infarct volume was calculated to compensate for the effect of brain edema. An edema index was calculated by dividing the total volume of the hemisphere ipsilateral to pMCAO by the total volume of the contralateral hemisphere. The actual infarct volume adjusted for edema was calculated by dividing the infarct volume by the edema index [37,45,54].

### 2.5. Evaluation of pyruvate action

In order to evaluate the effect of exogenous pyruvate administration on rat focal cerebral ischemia, three different doses of sodium pyruvate (250, 500 and 1000 mg/kg) were given to rats by intraperitoneal administration 30 min after the onset of pMCAO (n=9-13 animals per group). This treatment schedule and dosage range were based on a previous study in which pyruvate showed maximal neuroprotective effects in a rat model of global cerebral ischemia using these doses and treatment paradigm [23]. Rats were injected intraperitoneally with different volumes (0.75-3 mL) of a stock solution of sodium pyruvate (100 mg/mL) depending on the doses and body weight of the animals.

After investigating the dose-response relationship, we studied the effect of pyruvate (1000 mg/kg; i.p.) when administered 30 min before ischemia (n=26) and immediately after pMCAO (n=12). Moreover, we evaluated the effects of 500 and 1000 mg/kg of pyruvate when administered in a long-term administration regime (30 min, 6 h, 12 h and 18 h after pMCAO, n=13-14). In all cases pyruvate was dissolved in distilled water. As control, osmolarity-matched NaCl (209 mg/kg) solution was injected intraperitoneally. Results from our pilot studies indicated that there were no differences in infarct volume, mortality or neurological deficits among groups of rats that underwent pMCAO, when the vehicle was administered 30 min before, immediately after pMCAO, after 30 min of pMCAO or long-term administration.

*2.6. Data analysis*
Data are presented as means ± S.D. Values were compared using t-test, one way ANOVA with *post-hoc* Student-Newman-Keuls test and Mann-Whitney test for nonparametric variables (neurological scores). Differences were considered significant when $p<0.05$.

## 3. RESULTS

In this model of proximal pMCAO using an intraluminal nylon filament, TTC staining showed well-demarcated infarct areas in the temporoparietal cortex and in the laterocaudal part of the caudate putamen in all operated animals. High-grade neurological deficits (more than 2, see Methods section) were presented in all animals when tested at 24 h of pMCAO. Thus, no animals required exclusion on the basis of an inadequate degree of cerebral ischemia.

The effect of pyruvate on neurological deficits and mortality following pMCAO is shown in Table 1. Pyruvate administration at doses of 250 and 500 mg/kg after 30 min of pMCAO slightly reduced mortality and neurological deficits, but failed to significantly reduce infarct volume (Fig. 1). However, the highest dose of pyruvate (1000 mg/kg) did not reduce the neurological deficits or mortality compared with vehicle. On the contrary, treatment with this dose of pyruvate significantly increased total (cortical + subcortical) infarct volume compared to the vehicle group (Fig. 1). When considered separately, mean cortical infarct volume was increased by 27% by treatment with pyruvate when administered 30 min after pMCAO compared to vehicle rats (237.3 ± 23.9 and 186.8 ± 41.9 mm$^3$, respectively; p=0.025). Mean subcortical infarct volume was not modified by pyruvate treatment. Then, the significant increased in cortical infarct volume accounts for the significant increased in total brain infarct observed in pyruvate-treated animals.

The rostrocaudal distribution of cortical (Fig. 2A) and subcortical (Fig. 2B) infarct areas in the vehicle and pyruvate 1000 mg/kg groups is depicted in Fig. 2. Infarct areas were significantly greater (p<0.05) in pyruvate-treated rats than in the vehicle group at coronal levels 5 and 6, but in general there was a marked trend towards a greater cortical infarct areas in sections 2, 3 and 4, although these differences did not reach statistical significance (Fig. 2A). No changes in the rostrocaudal distribution of subcortical infarct areas were observed between vehicle and pyruvate-treated groups (Fig. 2B).

On the other hand, when pyruvate was administered 30 min before or immediately after pMCAO, no significant effect was noticed in infarct volume (Table 2). However, a significant increased in neurological deficits in pyruvate-pretreated rats as compared with vehicle was observed as shown in Table 1. Long-term administration of pyruvate failed to modify any of the evaluated parameters when compared to vehicle-treated rats (Tables 1 and 2).

The effects observed with pyruvate in the present study were not related to modification of physiological variables since these parameters (mean arterial blood pressure, $pO_2$, $pCO_2$, blood pH, rectal temperature, plasma glucose, hematocrit) were monitored in a separate experiment and did not differ between pyruvate-treated and vehicle-treated animals (data not shown). These findings are in agreement with those obtained by others [33,34], suggesting that exogenous administration of pyruvate does not significantly change physiological variables.

## 4. DISCUSSION

The core findings of this study are: i) pyruvate administration failed to confer protection against permanent focal cerebral ischemia in rats and ii) the highest dose of pyruvate increased infarct volume in rats subjected to pMCAO when treatment is given 30 min after the onset of ischemia.

This study was prompted by the encouraging results obtained by Lee et al [23], which showed that systemic administration of sodium pyruvate (500-1000 mg/kg) was remarkably neuroprotective in rats against global cerebral ischemia, a type of injury that mimics the clinical situation of cardio-respiratory arrest.

We decided to explore the effects of pyruvate at the exact dose range and similar treatment schedule of those tested by Lee et al [23] in rats subjected to pMCAO, because most cases of human ischemic stroke are caused by permanent occlusion of cerebral arteries. Since in stroke patients a very early spontaneous recanalization of an obstructed brain vessel is only rarely found, experimental models of pMCAO may be more relevant to the clinical situation [13,15,20,40].

Apparent discrepancies between our present results and those of Lee et al [23] may be due to differences in the pathophysiological mechanisms between the two models of cerebral ischemia. It is important to emphasize that in global cerebral ischemia, delayed neuronal death occurs in selective vulnerable regions of brain, specifically in CA1 region of hippocampus through a myriad of biochemical mechanisms that predominantly lead to apoptosis of damaged neurons [17,44,57]. In pMCAO models, most of the ischemic tissue dies through a rapid necrotic mechanism, which is accompanied by a dramatic inflammatory response [19,25,58]. Probably, the mechanism of neuronal death prevailing in each model is playing a key role, since pyruvate has been proven to limit apoptotic cell death in both non-neuronal cells [43] and in hippocampal and cortical neurons following forebrain ischemia, but does not reduce necrotic neuronal death induced by a 24-h exposure to NMDA, glutamate or ionomycin (calcium-overload toxicity) [23].

According to our results, the lowest doses of pyruvate reduced mortality and neurological deficits, but this favorable effects were not accompanied by a significant reduction in infarct volume (Table 1 and Fig. 1). This might reflect the fact that unlike ischemic injury to many other tissues, the severity of disability is not predicted well by the amount of brain tissue lost. For example, damage to a small area in the medial temporal lobe may lead to severe disability, while damage to a greater volume elsewhere has little effect of function [10]. The majority of studies directed toward determining neuroprotective efficacy have used reduction of infarct volume as a measure of a drug's efficacy in animals subjected to focal ischemia. Although it is presumed that reduced lesion size will translate to improved functional outcome, a direct correlation is not always observed in animals models [18] or in stroke patients [52]. For that reason, it is very important to emphasize that even when the lowest doses of pyruvate did not reduce infarct volume in pMCAO, one can not minimize the beneficial effects of these doses since a significant reduction in mortality (by 80 %) and improvement of neurological deficits were observed in the present study. Thus, further studies would be required to better characterize the effects of the lowest doses of pyruvate in models of cerebral ischemia to increase predictive outcome in the clinic.

On the other hand, there seems to be a threshold for the detrimental effects observed with pyruvate in focal ischemia and the time of administration also plays an important role. When the dose of pyruvate was increased to 1000 mg/kg (given 30 min after pMCAO), infarct volume was significantly increased by 27 %. Similarly, the detrimental effects of pyruvate in rats subjected to pMCAO are also observed when this glycolytic metabolite is administered before or immediately after pMCAO (Table 1), although no significant changes were noticed in infarct volumes (Table 2).

These observations suggest that high concentrations of pyruvate could counteract its neuroprotective effects probably by inducing an intracellular acidification. Pyruvate, as lactate, is transported across the plasma membrane by the $H^+$-monocarboxylate cotransporter [42], resulting in a cytosolic acidification [8]. The intracellular acidification is able to enhance the neurotoxic effect of $H_2O_2$ [8] and to induce the release of active iron from ferritin [1,14], a process that leads to enhanced production of hydroxyl radicals [46,51].

Most reports showing protective effects of pyruvate against cellular damage in different neuronal and non-neuronal cells, explain the protective properties of this metabolite through a mechanism involving $H_2O_2$ scavenging ability and not to an improvement of energy metabolism [8,30,31,43]. Unlike transient cerebral ischemia, oxidative damaging events do not seem to play a determining role in brain injury in permanent (no reperfusion) models of stroke [28]. This probably helps to explain our present results with pyruvate in a model of pMCAO.

In summary, the present study has evaluated by the first time the effects of pyruvate in permanent focal cerebral ischemia showing modest positive effects at low doses and detrimental effects when given at high doses. On the contrary of what was found in the case of transient global ischemia [23], present findings do not support a great neuroprotective role for pyruvate in permanent focal ischemia. We believe that it is very important to perform thorough, multifactorial and well-designed pre-clinical studies before assuming definitive conclusions on the neuroprotective effect of a given compound. In this particular case, our results and those by Lee et al [23] suggest that pyruvate could be tested in clinical trials with patients suffering from global cerebral damage but not in those with permanent stroke. Although success in animal studies does not guarantee success in clinical trials, the absence of neuroprotection or modest positive effects in animal studies indicate a lower likelihood of success in humans.

**Table 1.** Summary of mortality and neurological score of rats subjected to permanent middle cerebral artery occlusion (pMCAO) and effects of sodium pyruvate.

| Groups | Mortality | Neurological Score |
|---|---|---|
| Vehicle | 12 of 28 (42.8%) | 3.84 ± 1.02 |
| Pyruvate 250 mg/kg 30 min after pMCAO | 0 of 9 (0%) | 3.11 ± 0.73 |
| Pyruvate 500 mg/kg 30 min after pMCAO | 1 of 12 (8.3%) | 2.83 ± 0.79 * |
| Pyruvate 1000 mg/kg 30 min after pMCAO | 5 of 13 (38.5%) | 3.76 ± 1.12 |
| Pyruvate 1000 mg/kg Immediately after pMCAO | 4 of 12 (33.3%) | 4.3 ± 0.46 |
| Pyruvate 1000 mg/kg 30 min before pMCAO | 17 of 26 (65.4%) | 4.6 ± 0.63 ** |
| *Repeated treatments* | | |
| Pyruvate 500 mg/kg Repeated doses starting 30 min after pMCAO | 2 of 13 (15 %) | 3.47 ± 0.96 |
| Pyruvate 1000 mg/kg Repeated doses starting 30 min after pMCAO | 8 of 14 (57%) | 4.14 ± 1.06 |

A total of 127 adult rats were used for experiments. Neurological evaluation was performed as previously described [27,32] before sacrificing the animals at 24 h after pMCAO. $^{*}P<0.05$ and $^{**}P<0.01$ with respect to vehicle.

**Table 2.** Total infarct volumes of rats subjected to permanent middle cerebral artery occlusion (pMCAO) and treated with sodium pyruvate (1000 mg/kg) at the indicated time.

| **Groups** | **Infarct Volume (mm$^3$)** |
|---|---|
| Vehicle | 256.65 ± 48.77 |
| Pyruvate 1000 mg/kg, immediately after pMCAO | 275.12 ± 46.69 |
| Pyruvate 1000 mg/kg, 30 min before pMCAO | 305.9 ± 46.43 |
| *Repeated treatments* | |
| Pyruvate 500 mg/kg, repeated doses | 268.34 ± 45.32 |
| Pyruvate 1000 mg/kg, repeated doses | 271.05 ± 59.24 |

Values are mean ± S.D. for all groups. There were no statistically significant differences among groups.

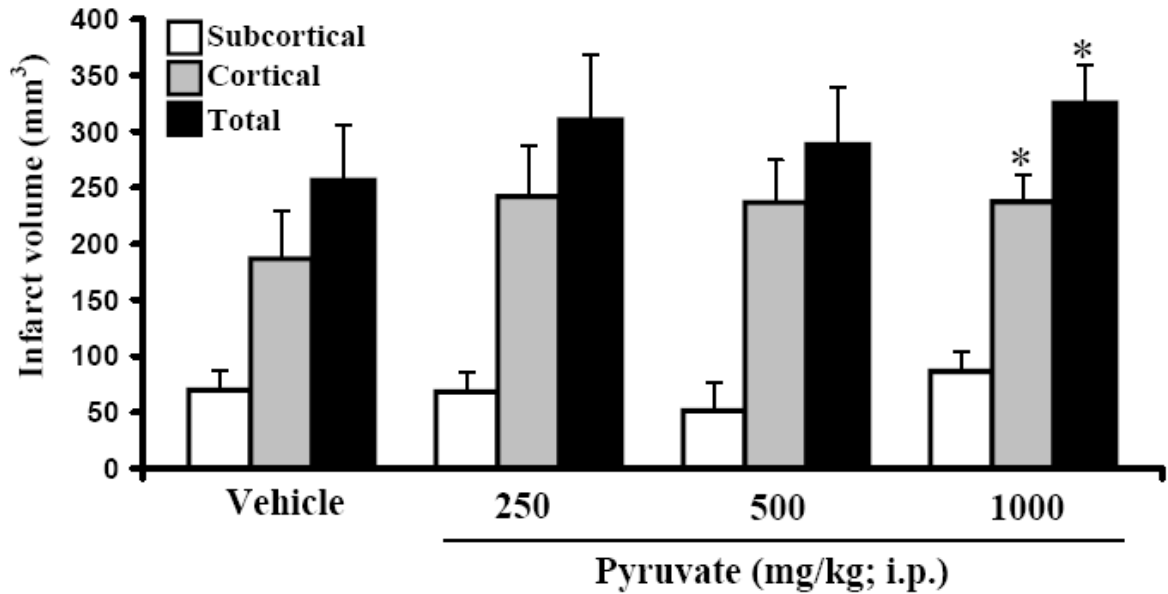

**Fig. 1.** Total, cortical and subcortical infarct volumes after permanent focal cerebral ischemia in rats. The animals received either pyruvate or vehicle 30 min after the onset of ischemia and were euthanized 24 h after stroke. There was a significant increase ($^*P<0.05$) in total and cortical volumes in the group treated with pyruvate 1000 mg/kg.

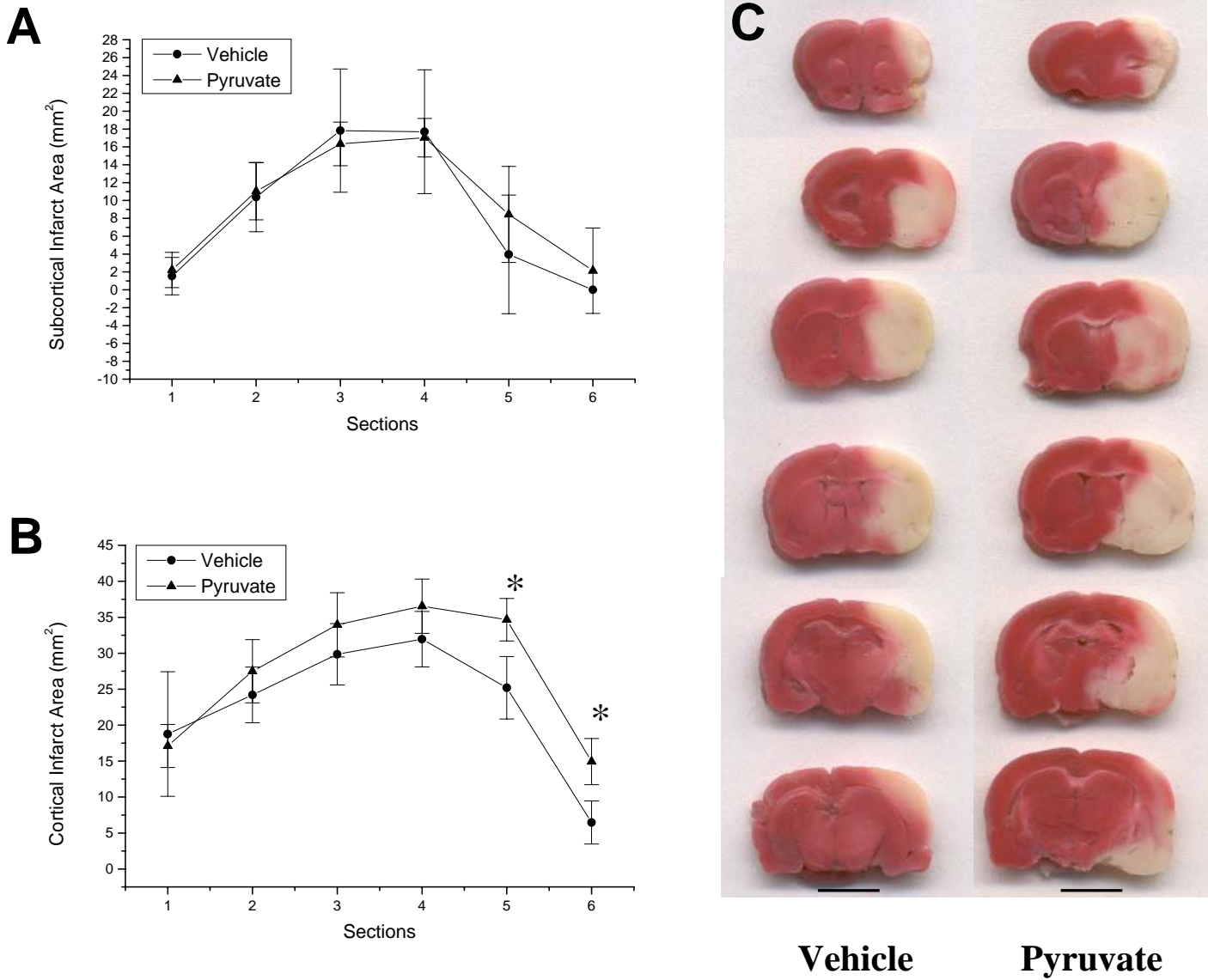

**Fig. 2**. Rostrocaudal distribution of areas of subcortical (**A**) and cortical infarction (**B**) at six coronal levels in pyruvate-treated (1000 mg/kg; 30 min after stroke) and vehicle-treated rats. Panel **C**: Representative 2,3,5-triphenyltetrazolium chloride-stained sections of vehicle and pyruvate-treated animals (1000 mg/kg; i.p.; 30 min after permanent middle cerebral artery occlusion). In panel **B**, $^{*}P<0.05$ compared with the vehicle. Bar= 1 cm in panel **C**.